\documentclass[twocolumn,showpacs,preprintnumbers,amsmath,amssymb]{revtex4}

\usepackage{graphicx}
\usepackage{dcolumn} 
\usepackage{bm}      
\usepackage{times}
\usepackage{color}

\newcommand{\vB}{\mathbf{B}}

\newcommand{\vF}{\mathbf{F}}
\newcommand{\vv}{\mathbf{v}}

\def\add#1{{#1}}

\begin{document}

\title{Cancellation exponent and multifractal structure in two-dimensional
       magnetohydrodynamics: direct numerical simulations and Lagrangian 
       averaged modeling}

\author{Jonathan Pietarila Graham, Pablo D. Mininni, and Annick Pouquet}
\affiliation{National Center for Atmospheric Research, P.O. Box 3000, 
             Boulder, Colorado 80307}

\date{\today}

\begin{abstract}
We present direct numerical simulations and Lagrangian averaged 
(also known as $\alpha$-model) simulations of forced and free 
decaying magnetohydrodynamic turbulence in two dimensions. 
The statistics of sign cancellations of the current at small 
scales is studied using both the cancellation exponent and the 
fractal dimension of the structures. The $\alpha$-model is 
found to have the same scaling behavior between positive and 
negative contributions as the direct numerical simulations. 
The $\alpha$ model is also able to reproduce the time 
evolution of these quantities in free decaying turbulence. At 
large Reynolds numbers, an independence of the cancellation 
exponent with the Reynolds numbers is observed.
\end{abstract}

\maketitle

The magnetohydrodynamic (MHD) approximation is often used 
to model plasmas or conducting fluids in astrophysical and 
geophysical environments. However, given the huge amount of temporal 
and spatial scales involved in the dynamics of these objects, simulations 
are always carried out in a region of parameter space far from the observed 
values. Lagrangian averaged magnetohydrodynamics (LAMHD), also called 
the MHD alpha-model \cite{Holm02,Mininni05} (or 
the Camassa-Holm equations in early papers studying the 
hydrodynamic case \cite{LANS}), has been recently introduced as a way 
to reduce the number of degrees of freedom of the system, while 
keeping accurate evolution for the large scales. This approach (as 
well as large eddy simulations, or LES, for MHD; see e.g. 
\cite{Agullo01}) is intended to model astrophysical or geophysical 
flows at high Reynolds numbers using available computational resources. 
Several aspects of the MHD alpha-model have already been tested in two 
and three dimensions at moderate Reynolds numbers, against direct 
numerical simulations of the MHD equations \cite{Mininni05}. 
These studies were focused on comparisons of the evolution of global 
quantities and the dynamics of the large scale components of the 
energy spectrum \cite{Mininni05,Ponty05}.

All these models introduce changes in the small scales 
in order to preserve the evolution of the large scales. 
In several cases, it is of interest to know the statistics 
of the small scales. It is also important to model properly the 
small scales because they have an effect on large scales, as for 
example in the case of eddy noise: the beating of two small scales 
eddies produces energy at the large scale, and this may affect the 
global long-time evolution of the flow, an issue that arises in 
global climate evolution or in solar-terrestrial interactions. 
Moreover, plasmas and conducting fluids generate thin and intense 
current sheets where magnetic reconnection takes place. In these 
regions, the magnetic field and the current rapidly change sign, 
and after reconnection the magnetic energy is turned into mechanical 
and thermal energy. These events are known to take place in the 
magnetopause \cite{Sonnerup81}, the magnetotail \cite{Birn96}, the 
solar atmosphere \cite{Gosling95}, and the interplanetary medium 
\cite{Schmidt03}.

Current sheets are strongly localized and intermittent. To preserve 
reliable statistics of these events in models of MHD turbulence is 
of utmost importance to model some of these astrophysical and geophysical 
problems. In this work, we study whether the MHD alpha-model is able 
to reproduce the statistics and scaling observed in these phenomena.

In order to measure fast oscillations in sign of a field on arbitrary 
small scales, the cancellation exponent was introduced 
\cite{Ott92,Vainshtein94,Sorriso02}. The exponent is a measure 
of sign-singularity. We can define the signed measure for the current 
$j_z({\bf x})$ on a set $Q(L)$ of size $L$ as
\begin{equation}
\mu_i(l) = \int_{Q_i(l)}{d{\bf x} \, j_z({\bf x})} \, \bigg{/} 
    \int_{Q(L)}{d{\bf x}\, |j_z({\bf x})|}
\label{eq:mu}
\end{equation}
where $\{Q_i(l)\}\subset Q(L)$ is a hierarchy of disjoint subsets 
of size $l$ covering $Q(L)$.  The partition function $\chi$ measures 
the cancellations at a given lengthscale $l$,
\begin{equation}
\chi(l) = \sum_{Q_i(l)}|\mu_i(l)|.
\end{equation}
\add{Note that for noninteger $L/l$ the subsets will not cover $Q(L)$ and
finite size box effects must be considered in the normalization
of Eq. (\ref{eq:mu}).}
We can study the scaling behaviors of the cancellations 
defining the cancellation exponent $\kappa$, where
\begin{equation}
\chi(l) \sim l^{-\kappa} .
\label{eq:scaling}
\end{equation}
Positive $\kappa$ indicates fast changes in sign on 
small scales (in practice, a cut-off is always present at the 
dissipation scale). A totally smooth field has $\kappa = 0$. This 
exponent can also be related with the fractal dimension $D$ of the 
structures \cite{Sorriso02},
\begin{equation}
\kappa = (d-D)/2 ,
\end{equation}
where $d$ is the number of spatial dimensions of the system. In some 
circumstances, we will also be interested on the cancellation exponent 
for the vorticity $\omega_z$. In that case the vorticity replaces the 
current in the definition of 
$\mu_i(l)$ [Eq. (\ref{eq:mu})]. 

Under special assumptions, relations between the cancellation exponent 
and scaling exponents have also been derived \cite{Vainshtein94}. 
Positive cancellation exponent $\kappa$ has been found in plasma 
experiments \cite{Ott92}, direct simulations of MHD turbulence 
\cite{Sorriso02}, in situ solar wind observations \cite{Carbone97}, 
and solar photospheric active regions \cite{Sorriso03}, where changes 
in the scaling were identified as preludes to flares.

In this work we will consider both free decaying and forced simulations 
of incompressible MHD and LAMHD turbulence in two dimensions (2D). 
The MHD equations in 2D can be written in terms of the stream function 
$\Psi$ and the $z$ component of the vector potential $A_z$,
\begin{eqnarray}
\partial_t \nabla^2 \Psi &=& [\Psi,\nabla^2\Psi] - [A_z,\nabla^2A_z] +
    \nu \nabla^4 \Psi \label{2DMHDmom} \\
\partial_t A_{z} &=& [\Psi,A_{z}] + \eta \nabla^2 A_z, \label{2DMHDind}
\end{eqnarray}
where the velocity and magnetic field are given by 
$\vv = \nabla \times (\Psi \hat{z})$ and 
$\vB = \nabla \times (A_z \hat{z})$ respectively, and 
$[F,G]= \partial_xF\partial_yG - \partial_xG\partial_yF$ is the 
standard Poisson bracket. The LAMHD equations are obtained by 
introducing a smoothing length $\alpha$, and the relation between 
smoothed (denoted by a subindex $s$) and unsmoothed fields is given 
by $\vF = (1-\alpha^2 \nabla^2) \vF_s$, for any field $\vF$. The system 
of LAMHD equations in this geometry \cite{Mininni05} is
\begin{eqnarray}
\partial_t \nabla^2 \Psi &=& [\Psi_s,\nabla^2\Psi] - [A_{s_z},\nabla^2A_z] +
    \nu \nabla^4 \Psi \label{2DLAMHDmom} \\
\partial_t A_{s_z} &=& [\Psi_s,A_{s_z}] + \eta \nabla^2 A_z. 
    \label{2DLAMHDind}
\end{eqnarray}
For both systems of equations, the current is given by 
$j_z = - \nabla^2 A_z$, and the vorticity by $\omega_z = - \nabla^2 \Psi$. 
\add{In these equations and in all the following figures, all quantities will
be given in familiar Alfv\'enic dimensionless units.}
Equations (\ref{2DMHDmom}-\ref{2DLAMHDind}) are solved in a periodic box 
using a pseudospectral code as described in \cite{Mininni05}. The code 
implements the 2/3-rule for dealiasing, and the maximum wavenumber 
resolved is $k_{max} = N/3$, where $N$ is the linear resolution used 
in the simulation. All the fields are written in dimensionless units.

To characterize the oscillating behavior and sign singularities in
the flows obtained from the MHD and LAMHD simulations, we perform a 
signed measure analysis and compute the cancellation exponent $\kappa$ 
for the current and for the vorticity. Following Eq. (\ref{eq:scaling}), 
its value is obtained by fitting $\chi(l) = c (l/L)^{-\kappa}$ through 
the inertial range, where $L=2\pi$ is the length of the box, and $c$ 
is a constant. The lengthscales in the inertial range used for this 
fit are obtained studying the scaling of the third order structure 
function \cite{Politano98}.

We first present results for a forced MHD simulation with $1024^2$ grid 
points, with $\eta = \nu = 1.6 \times 10^{-4}$. Both the momentum and
the vector potential equations were forced. The external forces had 
random phases in the Fourier ring between $k=1$ and $k=2$, and a 
correlation time of $\Delta t = 5 \times 10^{-2}$. The system was 
evolved in time until reaching a turbulent steady state. The amplitude 
of the magnetic force averaged over space was held constant to $0.2$, 
and the amplitude of the mechanical force to $0.45$, in order to have 
the system close to equipartition. 
Two more 
simulations using the LAMHD system were carried out, with the same parameters 
as the MHD run but with resolutions of $512^2$ grid points 
($\alpha \approx 0.0117$), and $256^2$ grid points 
($\alpha \approx 0.0234$) respectively \add{(the choice $\alpha = 2/k_{max}$ is conventional \cite{Mininni05,LANS})}.
The Kolmogorov's kinetic and 
magnetic dissipation wavenumbers in the MHD run are 
$k_\nu \approx k_\eta \approx 332$; 
in all the LAMHD simulations 
these wavenumbers are larger than the largest resolved wavenumber 
$k_{max}$, by virtue of the model. Note that although it is common 
to reduce the spatial resolution even more in studies of the large 
scale components of the energy spectrum in LES of hydrodynamic 
turbulence, this cannot be done in this context since \add{wide energy spectra
and large amounts of spatial statistics}
are needed to properly compute the cancellation exponent 
(see e.g. \cite{Cerutti98} for a study of intermittency in LES).

\begin {figure}
\includegraphics[width=8.5cm]{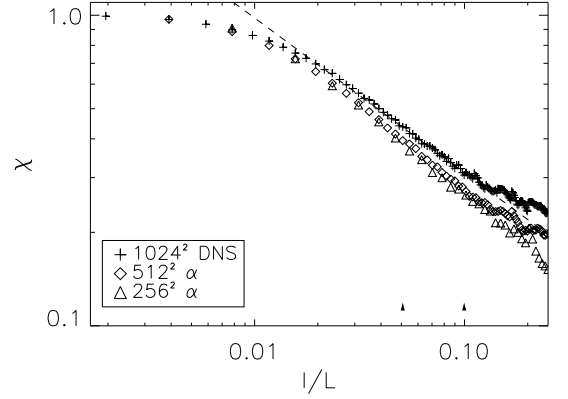}
\caption {$\chi(l)$ averaged in time for $j_z$ in forced MHD turbulence. 
     The pluses correspond to the $1024^2$ MHD simulation, diamonds to 
     the $512^2$ LAMHD run, and triangles to the $256^2$ LAMHD run. 
     The dashed line indicates a slope of \add{$0.50$.  The arrows indicate
	the inertial range.  Note that the slopes are of import, not the
	offsets.}}
\label{fig:forced}
\end{figure}

Fig. \ref{fig:forced} shows $\chi(l)$ for the three simulations, 
averaged using 11 snapshots of the current covering a total time 
span of 20 turnover times in the turbulent steady state. A power 
law can be identified at intermediate scales, scales smaller than 
the forcing band but larger than the dissipation scale. Note that 
the two LAMHD simulations reproduce the same scaling as the MHD 
simulation. As a result, the sign singularity and fractal structure 
are both well captured in the inertial range although the alpha-model 
is known to give thicker structures at scales smaller than $\alpha$ 
due to the introduction of the smoothing length \cite{Chen99,Mininni05}. 
The best fit for the current $j_z$ using a power law in the inertial 
range gives \add{$\kappa = 0.50 \pm 0.17$ for the $1024^2$ MHD run, 
$\kappa = 0.55 \pm 0.19$ for the $512^2$ LAMHD simulation, 
and $\kappa = 0.55 \pm 0.43$ for the $256^2$ LAMHD simulation. Note 
that a value of $\kappa = 0.50$ in the MHD simulation gives a value 
of the fractal dimension $D = 1.00 \pm 0.34$, close to the codimension 
of 1 corresponding to current sheets in MHD turbulence. For the 
vorticity, the cancellation exponent is $\kappa = 0.73 \pm 0.16$ for \
the $1024^2$ MHD run, $\kappa = 0.74 \pm 0.32$ for the $512^2$ LAMHD 
simulation, and $\kappa = 0.80 \pm 0.32$ for the $256^2$ LAMHD simulation, 
giving a fractal dimension of $D = 0.54$} in the MHD simulation. The 
values obtained are compatible with the values of $\kappa = 0.43\pm0.06$ 
and $D = 1.14\pm0.12$ for the current, and $\kappa = 0.69\pm0.12$ 
and $D = 0.62\pm0.24$ for the vorticity obtained in Ref. \cite{Sorriso02} 
for forced direct numerical simulations of 2D MHD turbulence using a 
$1024^2$ spatial grid and $\eta = \nu = 8 \times 10^{-4}$. Given the 
good agreement between MHD and LAMHD simulations, in the following 
we will only refer to the cancellation exponent for the current 
density.

\begin {figure}
\includegraphics[width=8.5cm]{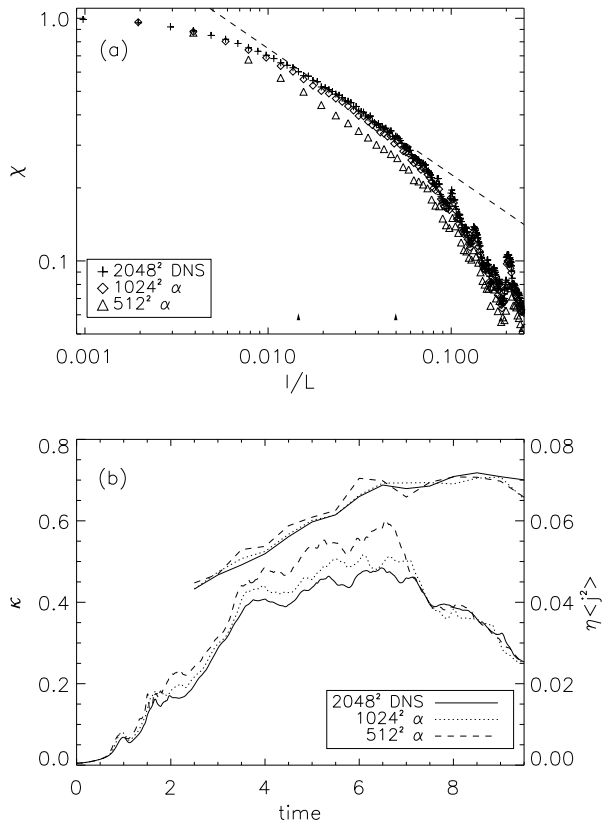}\\
\caption {(a) $\chi(l)$ at $t=4$ in the free decaying simulations, 
     pluses correspond to the $2048^2$ MHD simulation, diamonds to 
     the $1024^2$ LAMHD run, and triangles to the $512^2$ LAMHD run 
     (the dashed line indicates a slope of \add{$0.52$ and the arrows
	indicate the inertial range}); 
     (b) time history of the cancellation exponent (thick lines) for 
     the three runs, and of $\eta \left< j_z^2 \right>$, where the 
     brackets denote spatial average.}
\label{fig:decay}
\end{figure}

Fig. \ref{fig:decay}.a shows the corresponding results for free decaying 
MHD turbulence. Three simulations are shown, one MHD run using 
$2048^2$ grid points, a $1024^2$ LAMHD run with $\alpha \approx 0.0058 $, 
and a $512^2$ LAMHD run with $\alpha \approx 0.0117$. The three 
simulations were started with the same initial conditions; initial 
velocity and magnetic fields with random phases between $k=1$ and 
$k=3$ in Fourier space, and unit {\it r.m.s.} values. The kinematic 
viscosity and magnetic diffusivity used were $\nu = \eta = 10^{-4}$. 
The three simulations were evolved in time without external forces.

The evolution of the cancellation exponent as a function of time 
in the free decaying simulations is shown in Fig. \ref{fig:decay}.b. 
For these simulations, the cancellation exponent is computed between 
the lengthscales \add{$L/l \approx 20$ and $L/l \approx 70$}, where 
a power law scaling in $\chi(l)$ can be clearly identified \add{from
$t=2.5$} up to $t=10$. At $t = 0$ the cancellation exponent $\kappa$ 
is zero, which corresponds to the smooth initial conditions. 
\add{A gap between $t=0$ and $t=2.5$ is present where no clear scaling is
observed.}
As time evolves, $\kappa$ grows up to $0.75$ at $t \approx 8$, as the 
system evolves from the initially smooth fields to a turbulent state 
with strong and localized current sheets. After this maximum, the 
exponent $\kappa$ decays slowly in time. The maximum of $\kappa$ takes 
place slightly later than the maximum of magnetic dissipation, as is 
also shown in Fig. \ref{fig:decay}.b. Note that the alpha-model also 
captures the time evolution of the cancellation exponent in free 
decaying turbulence, as well as the fractal structure of the problem 
as time evolves.

As previously noted in \cite{Mininni05}, the alpha-model slightly 
overestimates the magnetic dissipation. Note however that in the 
three simulations the peak of magnetic dissipation takes place close 
to $t \approx 6$, just before the peak of the cancellation exponent 
$\kappa$. From the maximum energy dissipation rate, the Kolmogorov's 
dissipation wavenumber for the kinetic and magnetic energy at $t \approx 6$ 
are estimated as $k_\nu \approx k_\eta \approx 470$, and this is again 
larger than the largest wavenumbers resolved in the two LAMHD simulations.

The observed slow decay of the cancellation exponent (compared 
with the square current) is related to the persistence of strong 
current sheets in the system for long times, even after the peak of 
magnetic dissipation. 
The system, instead of evolving fast 
to a smooth solution at every point in space, keeps dissipating energy 
in a few thin localized structures. The existence of these current 
sheets at late times can be more easily verified in simulations with 
smaller viscosity $\nu$ and diffusivity $\eta$. While in the peak of 
magnetic dissipation the system is permeated by a large number of small 
current sheets, at late times only a few current sheets are observed 
isolated by large regions where the fields are 
smooth. 

\begin {figure}
\includegraphics[width=8.5cm]{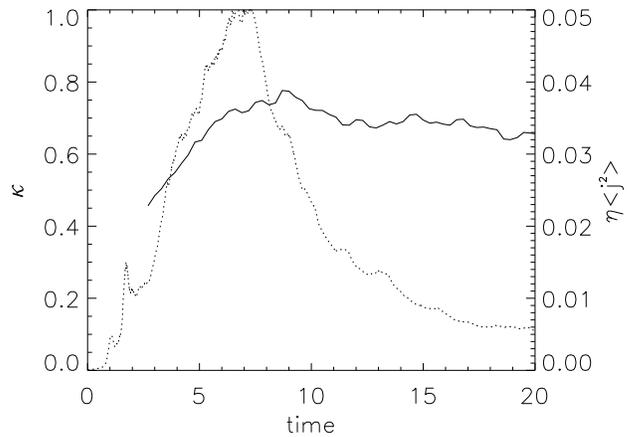}
\caption {Time history of $\kappa$ (solid line) and 
    $\eta \left< j_z^2 \right>$ (dotted line), for a free decaying 
    LAMHD simulation with $\eta=\nu=2 \times 10^{-5}$.}
\label{fig:reynl}
\end{figure}

Given the good agreement between direct numerical simulations (DNS) 
and LAMHD as seen in the preceding figure, we can reliably explore 
with the model Reynolds numbers unattainable in a reasonable time with 
DNS. In this context, we show that the maximum values of $\kappa$ 
obtained in the simulations seem to be insensitive to the Reynolds 
numbers within a given method (MHD or LAMHD) once a turbulent state 
is reached. As an example, in Fig. \ref{fig:reynl} we give the time 
history of the cancellation exponent and the square current for a 
free decaying LAMHD simulation with $\eta=\nu=2 \times 10^{-5}$ up 
to $t=20$. The initial conditions are the same as in the previously 
discussed simulations, and $\alpha \approx 0.0033$. It is worth noting 
that the time evolution of the magnetic dissipation in both decaying 
runs (Figs. \ref{fig:decay}.b and \ref{fig:reynl}) confirm previous 
results at lower Reynolds numbers \cite{Biskamp89,Politano89}: namely 
that the peak dissipation ($t \sim 7$) is lower for higher Reynolds 
numbers, while for later times it is quite independent of the Reynolds 
values.

\begin {figure}
\includegraphics[width=8.5cm]{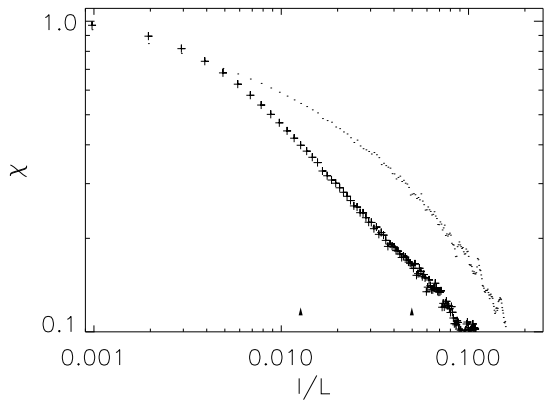}
\caption {$\chi(l)$ at \add{$t=3$} (dots), and $t=20$ (pluses), for the 
    free decaying LAMHD simulation with $\eta=\nu=2 \times 10^{-5}$. 
    The arrows indicate the inertial range.}
\label{fig:reync}
\end{figure}

Fig. \ref{fig:reync} shows $\chi(l)$ for early and late 
times in the same simulation. At small scales, the slope of 
$\chi$ always goes to zero, as can be expected since close to the 
dissipation lengthscale the fields are expected to be smooth. 
However, note that as time evolves the scaling of $\chi$ with 
$l$ drifts to smaller scales, and at $t=20$ a 
scaling can 
be observed up to $l/L \approx 0.005$.  \add{By virtue of the model the scaling
is wider and the slope goes to zero faster than in
the DNS due to the larger Reynolds number.}

The statistics of sign cancellation in magnetofluid turbulence are 
related with intermittency and anomalous scaling of structure 
functions \cite{Vainshtein94}, inherently associated with the 
dynamics of the small scales, and as a result harder to model 
in truncations or closures of the MHD equations. For example, 
two-point closures of turbulence behave smoothly (they also have 
no information about physical structures since they deal only with 
energy spectra). The intermittency of LES is an open topic, in 
particular because for neutral fluids the need to study the 
three dimensional case implied until recently that only low-resolution 
studies could be accomplished for which intermittent structures were 
barely resolved (see \cite{Kang03} for a recent study). From that 
point of view, the present study in two dimensions allows for higher 
Reynolds number studies. In MHD turbulence, the energy cascade being to 
small scales both in two and three space dimensions, it is hoped that the 
information gained here will carry on to the three-dimensional case. 
The new result stemming from this study is that the LAMHD alpha-model, 
although it alters the small scales through filtering, it nevertheless 
preserves some statistical information concerning the small scales. 
It is able to reproduce the scaling observed in forced MHD turbulence, 
as well as the time evolution of the cancellation exponent in free 
decaying simulations and as such, it represents a valuable model 
for studies of MHD flows for example at low magnetic Prandtl number 
$\nu/\eta$ as encountered in the liquid core of the Earth or in the 
solar convection zone (see e.g., \cite{Ponty05}).

\begin{acknowledgments}
Computer time provided by NCAR. NSF grant CMG-0327888 
at NCAR is gratefully acknowledged.
\end{acknowledgments}



\begin{thebibliography}{1}

\bibitem{Holm02}
D.D. Holm, Phys. D {\bf 170}, 253 (2002);
D.D. Holm, Chaos {\bf 12}, 518 (2002).

\bibitem{Mininni05}
P.D. Mininni, D.C. Montgomery, and A. Pouquet, Phys. Fluids {\bf 17}, 
035112 (2005);
P.D. Mininni, D.C. Montgomery, and A. Pouquet, Phys. Rev. E {\bf 71}, 
046304 (2005).

\bibitem{LANS}
D.D. Holm, J.E. Marsden and T.S. Ratiu, Adv. in Math. {\bf 137}, 1 (1998); 
S.Y. Chen, D.D. Holm, C. Foias, E.J. Olson, E.S. Titi, and S. Wynne, 
Phys. Rev. Lett. {\bf 81}, 5338 (1988); S.Y. Chen, C. Foias, D.D. Holm, 
E. Olson, E.S. Titi, and S. Wynne, Physica D {\bf 133} 49 (1999); 
S.Y. Chen, C. Foias, D.D. Holm, E.J. Olson, E.S. Titi, and S. Wynne, 
Phys.  Fluids {\bf 11}, 2343 (1999).

\bibitem{Agullo01}
O. Agullo, W.-C. M\"uller, B. Knaepen, and D. Carati, Phys. Plasmas 
{\bf 8}, 3502 (2001).

\bibitem{Ponty05}
Y. Ponty, P.D. Mininni, D.C. Montgomery, J.-F. Pinton, H. Politano, 
and A. Pouquet, Phys. Rev. Lett. {\bf 94}, 164502 (2005).

\bibitem{Sonnerup81}
B.U.O. Sonnerup, G. Paschmann, I. Papamastorakis, N. Sckopke, G. 
Haerendel, S.J. Bame, J.R. Asbridge, J.T. Gosling, and C.T. Russell, 
J. Geophys. Res. {\bf 86}, 10049 (1981).

\bibitem{Birn96}
J. Birn and M. Hesse, J. Geophys. Res. {\bf 101}, 15345 (1996).

\bibitem{Gosling95}
J.T. Gosling, J. Birn, M. Hesse, Geophys. Res. Lett. {\bf 22}, 869 (1995).

\bibitem{Schmidt03}
J.M. Schmidt, and P.J. Cargill, J. Geophys. Res. {\bf 108}, 1023 (2003).

\bibitem{Ott92}
E. Ott, Y. Du, K.R. Sreenivasan, A. Juneja, and A.K. Suri, 
Phys. Rev. Lett. {\bf 69}, 2654 (1992).

\bibitem{Vainshtein94}
S.I. Vainshtein, K.R. Sreenivasan, R.T. Pierrehumbert, V. Kashyap, and 
A. Juneja, Phys. Rev. E {\bf 50}, 1823 (1994).

\bibitem{Sorriso02}
L. Sorriso-Valvo, V. Carbone, A. Noullez, H. Politano, A. Pouquet, and 
P. Veltri, Phys. Plasmas {\bf 9}, 89 (2002).

\bibitem{Carbone97}
V. Carbone and R. Bruno, Astroph. J. {\bf 488}, 482 (1997).

\bibitem{Sorriso03}
L. Sorriso-Valvo, V. Carbone, V. Abramenko, V. Yurchysshyn,
A. Noullez, H. Politano, A. Pouquet, and P.L. Veltri, Planet. 
Space Sc. {\bf 52}, 937 (2004).

\bibitem{Politano98}
H. Politano and A. Pouquet, Phys. Rev. E {\bf 57}, R21 (1998); 
H. Politano and A. Pouquet, Geophys. Res. Lett. {\bf 25} 273 (1998).

\bibitem{Cerutti98}
S. Cerutti and C. Meneveau, Phys. Fluids {\bf 10}, 928 (1998).

\bibitem{Chen99}
S.Y. Chen, D.D. Holm, L.G. Margolin, and R. Zhang, Phys. D {\bf 133}, 
66 (1999).

\bibitem{Biskamp89}
D. Biskamp and H.Welter, Phys. Fluids B {\bf 1}, 1964 (1989).

\bibitem{Politano89}
H. Politano, A. Pouquet, and P.L. Sulem, Phys. Fluids B {\bf 1}, 2330 (1989).

\bibitem{Kang03}
H.S. Kang, S. Chester, and C. Meneveau, J. Fluid. Mech. {\bf 480}, 
129 (2003).

\end{thebibliography}
\end{document}